\documentclass[aps,prd,groupedaddress,preprintnumbers,superscriptaddress,10.5pt,preprint,nofootinbib]{revtex4-1}

\usepackage{amssymb, amsfonts, amsmath, mathtools, bm}
\usepackage{color}

\usepackage{booktabs}
\usepackage[caption=false]{subfig}
\usepackage{mathrsfs}
\usepackage{enumerate}
\usepackage{amsfonts}
\usepackage{graphicx}
\usepackage{enumitem}
\usepackage{setspace}
\newcommand{\bra}[1]{\left( #1 \right)}
\newcommand{\brb}[1]{\left[ #1 \right]}
  
\newcommand{\be}{\begin{equation}}  
\newcommand{\ee}{\end{equation}}
\newcommand{\D}{{\rm d}}
\newcommand{\fr}[2]{\frac{#1}{#2}}

\newcommand{\ti}{\tilde}
\newcommand{\na}{\nabla}

\newcommand{\Mpl}{M_{\rm Pl}}
\newcommand{\mn}{{\mu \nu}}
\newcommand{\ep}{\epsilon}
\newcommand{\mL}{\mathcal{L}}
\newcommand{\rmd}{\mathrm{d}}
\newcommand{\eEFT}{\ep_{\rm EFT}}
\newcommand{\rH}{r_{\rm H}}

\let\originalleft\left
\let\originalright\right
\renewcommand{\left}{\mathopen{}\mathclose\bgroup\originalleft}
\renewcommand{\right}{\aftergroup\egroup\originalright}

\usepackage[
]{hyperref}
\hypersetup{%
 setpagesize=false,
 bookmarksnumbered=true,%
 bookmarksopen=true,%
 colorlinks=true,%
 linkcolor=blue,%
 citecolor=red}

\begin{document}

\title{
Probing higher curvature gravity via ringdown with overtones
}

\author{Keisuke Nakashi}
\affiliation{Kaichi Tokorozawa Secondary School, 169 Ohaza Matsugo, Tokorozawa City, 
Saitama 395-0015, Japan}
\affiliation{Department of Physics, Rikkyo University, Toshima, Tokyo 171-8501, Japan}
\author{Masashi Kimura}
\affiliation{Department of Information, Artificial Intelligence and Data Science,
Daiichi Institute of Technology, Tokyo 110-0005, Japan}
\affiliation{Department of Physics, Rikkyo University, Toshima, Tokyo 171-8501, Japan}
\author{Hayato Motohashi}
\affiliation{Department of Physics, Tokyo Metropolitan University, 1-1 Minami-Osawa, Hachioji, Tokyo 192-0397, Japan}
\affiliation{Yukawa Institute for Theoretical Physics, Kyoto University, 606-8502, Kyoto, Japan}
\author{Kazufumi Takahashi}
\affiliation{Department of Physics, College of Humanities and Sciences, Nihon University, Tokyo 156-8550, Japan}
\affiliation{Yukawa Institute for Theoretical Physics, Kyoto University, 606-8502, Kyoto, Japan}

\date{\today}
\preprint{RUP-25-27}

\begin{abstract}
We investigate metric perturbations of a spherically symmetric black hole in higher curvature gravity.
We show that higher curvature corrections deform the near-horizon region of the effective potential, and that the deviations of the quasinormal mode (QNM) frequencies from their general relativity (GR) values become more pronounced for overtone modes.
We find that, as the order of the higher curvature term increases, the deformations approach the horizon and the deviations of the overtone QNM frequencies grow progressively larger.
We also analyze the ringdown waveforms in the higher curvature gravity model.
We consider setups in which the deviations from the vacuum-GR QNMs remain mild for the fundamental mode and the first few overtones, and show that these shifted QNMs can be identified in the ringdown signal through waveform fitting.
\end{abstract}

\maketitle

\section{Introduction}

Since the detection of gravitational waves (GWs) from binary black hole coalescences~\cite{LIGOScientific:2016aoc,LIGOScientific:2018mvr,LIGOScientific:2020ibl,LIGOScientific:2021usb,KAGRA:2021vkt,LIGOScientific:2025slb}, we have been able to test general relativity (GR) in the regime where spacetime is extremely curved and dynamical~\cite{Barack:2018yly,LIGOScientific:2021sio,Berti:2025hly}. 
Such a development provides us with an opportunity to explore possible signatures of new physics. 

There are several motivations for considering gravitational theories that extend GR. 
One such motivation is the incompleteness of GR in the ultraviolet regime, and hence the development of theories that modify GR in the high-energy regime has been pursued so far (see Ref.~\cite{Berti:2015itd} and references therein for example). 
One natural possibility for such modifications in the high-energy regime is to incorporate higher derivative corrections into the Einstein-Hilbert action. 
For instance, string theory predicts that an infinite series of higher curvature correction terms are added to the Einstein-Hilbert action~\cite{Gross:1986mw,Gross:1986iv}.

Such correction terms are also motivated from the perspective of effective field theory (EFT), which is itself suggested by more fundamental physics. Furthermore, exploring which type of EFTs can be tested through GW observations is of significant importance, as it provides valuable feedback to fundamental physics. 
Given the wide range of possible extensions of GR, it is common to impose physically motivated assumptions such as the following:
the contributions from modifications to GR are characterized by an energy scale higher than the curvature scale around the compact object emitting GWs, and the contributions from higher curvature terms are assumed to be sufficiently small to be treated perturbatively. 
Furthermore, we assume that such EFTs should be testable through GW observations, be consistent with other experiments and observations, including short distance tests of GR, and not involve new light degrees of freedom.\footnote{For EFTs that include additional light degree(s) of freedom, see extensive studies in the context of scalar-tensor gravity~\cite{Arkani-Hamed:2003pdi,Arkani-Hamed:2003juy,Cheung:2007st,Gubitosi:2012hu,Franciolini:2018uyq,Mukohyama:2022enj,Mukohyama:2025jzk} and vector-tensor gravity~\cite{Cheng:2006us,Mukohyama:2006mm,Aoki:2021wew,Aoki:2023bmz}.}
The Lagrangian of the EFT obeying these assumptions has been constructed in Ref.~\cite{Endlich:2017tqa}, and its effects on black hole spacetimes have also been investigated~\cite{Cardoso:2018ptl,Cano:2019ore}. 

The GWs emitted immediately after the binary black hole merger, the so-called ringdown, are well suited for testing GR in the strong and dynamical regime of gravitational fields~\cite{Berti:2025hly}. 
In GR, it is well known that the ringdown can be well described by the quasinormal modes (QNMs) of the black hole~\cite{Buonanno:2006ui}, which
can be predicted using black hole perturbation theory~\cite{Leaver:1985ax, Leaver:1986gd}. 
According to the black hole uniqueness theorem in GR, the Kerr black hole is the unique stationary and asymptotically flat vacuum solution, and its mass and angular momentum completely determine the QNM frequencies. 
It is expected that testing gravity theories using the ringdown phase of GWs allows us to detect beyond GR effects. 
The actual observational data of GWs are consistent with the fundamental mode of QNMs predicted in GR~\cite{LIGOScientific:2021sio}. 
Meanwhile, it has been pointed out that including overtones in the fitting model is important for improving the agreement with the numerical waveform and extracting the information of the remnant black hole parameters~\cite{Baibhav:2017jhs,Bhagwat:2019dtm,Giesler:2019uxc,Cook:2020otn,Forteza:2021wfq,Baibhav:2023clw,Nee:2023osy,Takahashi:2023tkb}. 
There has been active debate on whether overtone information can be reliably extracted from GW observations. 
Some analyses claim that the observational data contain evidence of the overtone contributions~\cite{Isi:2019aib,Isi:2022mhy,Finch:2022ynt,LIGOScientific:2025wao,LIGOScientific:2025rid}, whereas others argue that such signals remain buried in the noise and cannot be robustly identified~\cite{Cotesta:2022pci,Carullo:2023gtf}.
The verification of gravity theories using overtones is anticipated to become increasingly important as the sensitivity of future GW detectors improves. 

In EFTs with higher curvature corrections, modifications to GR introduce corrections to the QNM frequencies. 
By comparing the QNM frequencies with the ringdown waveform, it becomes possible to explore the validity of GR and which EFTs are justified. 
Driven by this motivation, numerous studies have been conducted to calculate the corrections to the QNM frequencies of black holes in EFTs, focusing on both static spherically symmetric black holes~\cite{Cardoso:2018ptl,deRham:2020ejn,Hirano:2024fgp,Cardoso:2019mqo,Silva:2024ffz} and rotating black holes~\cite{Cano:2020cao,Cano:2021myl,Cano:2023tmv,Cano:2023jbk,Cano:2024ezp,Maenaut:2024oci}. 
Furthermore, there are some attempts to obtain constraints on EFTs from the observational GW data~\cite{Silva:2022srr,Liu:2024atc}. 
Complementary to such a theory-specific approach, one can directly parametrize the deformation of the effective potential in the GR master equation for GWs and examine how it affects the QNMs.
This approach, known as the parametrized QNM formalism~\cite{Cardoso:2018ptl,Hirano:2024fgp}, allows for model-independent predictions of QNM frequencies~\cite{Volkel:2022aca,Volkel:2022khh,Franchini:2022axs}.

In the present paper, we investigate how higher curvature corrections affect the QNMs and their imprint on the ringdown waveform. 
We focus on a class of higher curvature gravity theories that admit the Schwarzschild spacetime as a background solution while modifying the dynamics of perturbations. 
The higher curvature terms deform the effective potential in the near-horizon region, and the deformation approaches the horizon as the order of the higher curvature terms becomes larger. 
Using the parametrized QNM formalism, we first compute the frequency shifts of QNMs induced by the higher curvature corrections. 
In particular, we demonstrate that the overtone frequencies tend to deviate more strongly from their GR values as the deformation approaches the horizon, while the fundamental mode is less affected.
This is similar to the result of Ref.~\cite{Konoplya:2022pbc} for a different type of near-horizon deformation, where this phenomenon was referred to as overtone outbursts, as well as to observations in the context of spectral instability induced by deformations located far from the black hole~\cite{Barausse:2014tra,Jaramillo:2020tuu,Jaramillo:2021tmt,Cheung:2021bol}.
Although our results are valid only to linear order in the deviation from GR, this behavior suggests a connection between the higher curvature corrections and the overtone outbursts. 

To assess whether these QNM deviations manifest in observable signals, we compute time-domain waveforms by solving the master equation. 
The authors of Refs.~\cite{Thomopoulos:2025nuf,deMedeiros:2025ayq} computed time-domain waveforms assuming a small deviation from the Regge-Wheeler equation in GR, using parametrized potentials that contain a single power-law correction of the form~$(r_{\rm H}/r)^j$, with $r_{\rm H}$ the horizon radius of the background black hole, and focused on relatively small values of $j$.
In contrast, in the present paper we explore larger values of $j$, which were not extensively explored in the previous studies, and we take into account the contributions from the first as well as higher overtones.
Also, we perform ringdown fits using templates built from the EFT QNMs. 
By comparing fits that incorporate the EFT QNMs with those that use the GR QNMs, we show that the EFT-based templates achieve systematically better agreement, especially at early times when the overtones dominate the waveform. 
These results indicate that the excitation of EFT QNMs may provide observational access to near-horizon physics and can potentially enhance the sensitivity of black hole spectroscopy to higher curvature gravity. 

This paper is organized as follows. 
In Sec.~\ref{sec:model}, we briefly discuss the higher curvature gravity model we consider.
We also study odd-parity perturbations around the Schwarzschild solution in our model to derive a master equation whose effective potential is slightly deformed from the Regge-Wheeler potential in GR. 
In Sec.~\ref{sec:pQNM}, we review the parametrized QNM formalism and explain how the shifts of the QNM frequencies relate to the deformation of the effective potential. 
In particular, we show that the overtone frequencies tend to deviate increasingly from the GR values as the order of the higher curvature term becomes larger. 
In Sec.~\ref{sec:time-domain}, we obtain the time-domain waveform by solving the master equation numerically and perform a fitting analysis with a superposition of QNMs. 
Finally, we present our summary and discussions in Sec.~\ref{sec:summary}.

\section{Higher curvature gravity and perturbations}\label{sec:model}

\subsection{Theory and background solution}

To study the effect of higher curvature corrections on gravity theories, we consider the following action:
    \be
    S=\int \D^4x\sqrt{-g}\,\fr{\Mpl^2}{2}\brb{R-\fr{\mathtt{a}(C)\ti{C}^2}{\Lambda^6}},
    \label{action}
    \ee
where $\ti{C}\coloneqq \epsilon^{\alpha\beta}{}_{\mu\nu}R^{\mu\nu\gamma\delta}R_{\alpha\beta\gamma\delta}$, $\epsilon_{\alpha\beta \mu\nu}$ is the Levi-Civita tensor with $\epsilon_{0123} = \sqrt{-g}$, and $\Lambda$ is a constant of mass dimension one.
The scalar function~$\mathtt{a}(C)$, with $C\coloneqq R^{\alpha\beta\gamma\delta}R_{\alpha\beta\gamma\delta}$, is assumed to be dimensionless.
We mainly consider cases where $\mathtt{a}(C)$ is a polynomial in $C$ whose coefficients are sufficiently small to allow a perturbative treatment.
Since each term in the polynomial contributes independently at leading order in deviations from GR, we will focus on the representative form~$\mathtt{a}(C)\propto C^p$ with $p$ a non-negative integer. 
It should be noted that the second term in Eq.~\eqref{action} involves higher derivatives of the metric, which leads to the problem of Ostrogradsky ghost~\cite{Motohashi:2014opa,Woodard:2015zca,Motohashi:2020psc,Aoki:2020gfv} in principle, as demonstrated in stability analysis of black hole perturbations in theories involving $\ti{C}$~\cite{Motohashi:2011pw,Motohashi:2011ds}. 
Nevertheless, the theory would make sense as a low-energy EFT for perturbations of Ricci-flat spacetimes, including the Kerr spacetime.

Varying the action~\eqref{action} with respect to the metric, we obtain the field equation,
\begin{align}
R_{\mu \nu} - \frac{1}{2} g_{\mu \nu}R 
=
\frac{1}{\Lambda^6}
\left[
\frac{\tilde{C}^2}{2}g_{\mu \nu}(\mathtt{a}^\prime C +  \mathtt{a})
+
4 R_{\mu \alpha \nu \beta} \nabla^\alpha \nabla^\beta(\mathtt{a}^\prime \tilde{C}^2 )
+
8 \tilde{R}_{\mu \alpha \nu \beta}\nabla^\alpha \nabla^\beta(\mathtt{a} \tilde{C})
+{\cal E}_{\mu\nu}
\right],
\label{eomEFT}
\end{align}
where $\mathtt{a}^\prime \coloneqq \D \mathtt{a}/\D C$ and $\tilde{R}_{\alpha\beta\gamma\delta}\coloneqq \epsilon_{\alpha\beta\mu\nu}R^{\mu\nu}{}_{\gamma\delta}$.
Also, ${\cal E}_{\mu\nu}$ is a tensor that vanishes when evaluated on a Ricci-flat spacetime.
Equation~\eqref{eomEFT} can be used to study higher curvature corrections to Ricci-flat solutions in GR. When one is interested in the leading-order correction, the last term~${\cal E}_{\mu\nu}$ in Eq.~\eqref{eomEFT} can be neglected. 
For simplicity, in what follows, we consider a static and spherically symmetric metric as the background spacetime.
Since $\ti{C}$ vanishes due to the spherical symmetry, the background spacetime is described by the Schwarzschild metric,
    \begin{align}
    \begin{split}
    \bar{g}_\mn\D x^\mu\D x^\nu
    &=-f(r)\D t^2+\fr{\D r^2}{f(r)}+r^2 \gamma_{ab}\D x^a\D x^b,
    \\
    f(r) &\coloneqq 1-\fr{\rH}{r},
    \end{split}\label{eq:schwarzschild}
    \end{align}
with $\rH$ being a constant. 
Here, the indices~$a,b,\cdots$ denote the angular variables~$\{\theta,\varphi\}$, and $\gamma_{ab}\D x^a\D x^b=\D \theta^{2} + \sin^{2} \theta\, \D \varphi^{2}$ is the line element of a two-dimensional sphere.
Note that at the background level $C$ takes the nonzero value $C_{\rm BG}=12\rH^2/r^6$, whereas $\tilde{C}$ vanishes.

The theory described by the action~\eqref{action} is a generalization of the gravity theory with a $\ti{C}^2$ correction~\cite{Endlich:2017tqa, Cardoso:2018ptl}, which corresponds to $\mathtt{a}(C)=1$.
As in the $\ti{C}^2$-corrected theory, the model in Eq.~\eqref{action} not only admits the Schwarzschild metric as an exact solution but also possesses several useful structural features for studying linear perturbations:
\renewcommand{\theenumi}{(\roman{enumi})}
\renewcommand{\labelenumi}{\theenumi}
\begin{enumerate}[itemsep=-5pt, topsep=0pt]
\item At the level of linear perturbations, only the odd-parity modes exhibit deviations from GR, while the even-parity modes remain unaffected.
\item For a single monomial in $\mathtt{a}(C)$, the master equation for the odd-parity perturbation contains only a single correction term in the effective potential.
\item The higher the curvature order, the more closely the correction in the effective potential is localized to the horizon, as expected in higher curvature gravity theories.
\end{enumerate}
We shall discuss these points in detail in the next section.

\subsection{Gravitational perturbation}

Let us now study gravitational perturbations around the background spacetime~\eqref{eq:schwarzschild}.
In our system, the dynamics of even-parity perturbations is identical to the vacuum GR case at the linear level because $\ti{C}$ starts at second order in the even-parity perturbations. Therefore, we focus on the odd-parity perturbations, where the effects of the higher curvature corrections appear already at the linear level.
The perturbations~$h_\mn\coloneqq g_\mn-\bar{g}_\mn$ can be decomposed as follows:
    \be
	\begin{split}
	h_{tt}&=h_{tr}=h_{rr}=0,\\
	h_{ta}&=\sum_{\ell,m}r^2h_{0,\ell m}(t,r)E_a{}^b\bar{\na}_bY_{\ell m}(\theta,\varphi),\\
	h_{ra}&=\sum_{\ell,m}r^2h_{1,\ell m}(t,r)E_a{}^b\bar{\na}_bY_{\ell m}(\theta,\varphi),\\
	h_{ab}&=\sum_{\ell,m}r^2h_{2,\ell m}(t,r)E_{(a}{}^c\bar{\na}_{|c|}\bar{\na}_{b)}Y_{\ell m}(\theta,\varphi),
	\end{split} \label{pert_odd}
	\ee
where $Y_{\ell m}$ is the spherical harmonics, $E_{ab}$ is the completely antisymmetric tensor defined on a two-dimensional sphere, and $\bar{\na}_a$ denotes the covariant derivative with respect to $\gamma_{ab}$.
Note that we have inserted $r^2$ in front of $h_{0}$, $h_{1}$, and $h_{2}$ for later convenience.
The odd-parity part of an infinitesimal coordinate transformation~$x^\mu\to x^\mu+\ep^\mu$ can be written as
	\be
	\ep^t=\ep^r=0, \qquad
	\ep^a=\sum_{\ell,m}\Xi_{\ell m}(t,r)E^{ab}\bar{\na}_bY_{\ell m}(\theta,\varphi).
	\ee
Correspondingly, the gauge transformation of the coefficients $h_0$, $h_1$, and $h_2$ is given by
    \be
    h_0\rightarrow h_0-\partial_{t} \Xi, \quad
	h_1\rightarrow h_1- \partial_{r} \Xi, \quad
	h_2\rightarrow h_2-2\Xi.
    \label{gaugetrnsf_odd}
    \ee
This implies that the gauge condition~$h_2=0$ completely fixes the gauge degrees of freedom. 
In what follows, we impose this gauge condition at the action level~\cite{Motohashi:2016prk} and study the quadratic Lagrangian written in terms of the remaining variables~$h_0$ and $h_1$.
After a straightforward calculation, we obtain the following action:
	\be
	S^{(2)}_{\rm odd}=\int \D t \D r\, \mL^{(2)}_{\rm odd},
	\ee
where
	\be
	\fr{2\ell+1}{2\pi \ell(\ell+1)}\mL^{(2)}_{\rm odd}=
	a_1h_0^2-a_2h_1^2+a_3\bra{\partial_{t} h_1 - \partial_{r} h_0}^2. \label{qlag_odd}
	\ee
Here, the coefficients~$a_1$, $a_2$, and $a_3$ are given by
    \begin{gather}
    \begin{split}
    &a_1=\fr{\Mpl^2}{2}\fr{\brb{\ell(\ell+1)-2}r^2}{f}, \qquad
    a_2=\fr{\Mpl^2}{2}\brb{\ell(\ell+1)-2}r^2f,
    \\
    &a_3=\fr{\Mpl^2}{2}r^4\brb{1-\fr{1152\ell(\ell+1)\mathtt{a}(C_{\rm BG})}{(\rH\Lambda)^6}\bra{\fr{\rH}{r}}^8}.
    \label{p-coeffs}
    \end{split}
    \end{gather}
Note that only $a_3$ is subjected to the higher derivative correction.
We now introduce an auxiliary variable~$\chi$ to rewrite the Lagrangian~\eqref{qlag_odd} as
\begin{align}
	\fr{2\ell+1}{2\pi \ell(\ell+1)}\mL^{(2)}_{\rm odd}
    =
	a_1h_0^2-a_2h_1^2
    +a_3\bra{\partial_{t} h_1 - \partial_{r} h_0}^2
    -a_3\brb{\chi-\bra{\partial_{t} h_1 - \partial_{r} h_0}}^2. \label{qlag_odd2}
\end{align}
One can easily see that this Lagrangian is equivalent to Eq.~\eqref{qlag_odd}.
In fact, the equation of motion for $\chi$ yields $\chi=\partial_{t} h_1 - \partial_{r} h_0$, and then substituting this back into Eq.~\eqref{qlag_odd2} recovers the original Lagrangian~\eqref{qlag_odd}.
Note that the last term in Eq.~\eqref{qlag_odd2} has been introduced so that it removes the terms quadratic in $\partial_{r}h_0$ and $\partial_{t}h_1$ from the Lagrangian.
As a result, the equations of motion for $h_0$ and $h_1$ obtained from Eq.~\eqref{qlag_odd2} yield
    \be
    h_0=-\fr{\partial_{r}(a_3\chi)}{a_1}, \qquad
    h_1=-\fr{a_3\partial_{t}\chi}{a_2}.
    \ee
Substituting these back into Eq.~\eqref{qlag_odd2}, we obtain the following Lagrangian written in terms of a single master variable:
\begin{align}
	\fr{2\ell+1}{2\pi \ell(\ell+1)}\mL^{(2)}_{\rm odd}=
	\fr{a_3^2}{a_2}(\partial_{t}\chi)^2-\fr{a_3^2}{a_1}(\partial_{r}\chi)^2+a_3\brb{\partial_{r}\bra{\fr{\partial_{r}a_3}{a_1}}-1}\chi^2. \label{qlag_odd3}
\end{align}
Finally, by introducing the tortoise coordinate~$r_*$ such that
    \be
    \fr{\D r_*}{\D r}=\sqrt{\fr{a_1}{a_2}}
    =\fr{1}{f},
    \ee
and a new master variable
    \be
    \Psi\coloneqq \fr{a_3}{(a_1a_2)^{1/4}} \chi,
    \ee
the action takes the form
	\be
	S^{(2)}_{\rm odd}=\int \D t\D r_*\, \ti{\mL}^{(2)}_{\rm odd},
	\ee
where $\ti{\mL}^{(2)}_{\rm odd}=\sqrt{a_2/a_1}\,\mL^{(2)}_{\rm odd}$ and
	\be
	\ti{\mL}^{(2)}_{\rm odd}
    \propto \fr{1}{2} \bra{\partial_{t} \Psi}^2-\fr{1}{2}\bra{\partial_{r_*}\!\Psi}^2-\fr{1}{2}V_{\rm eff}\Psi^2.
    \ee
Here, we have omitted an irrelevant overall constant.
From this Lagrangian, we obtain the following equation of motion:   
    \be
    \partial_{t}^{2}{\Psi} - \partial_{r_*}^{2}\!\Psi+V_{\rm eff}\Psi=0. \label{RWeq}
    \ee
The effective potential~$V_{\rm eff}$ can be written as
    \be
    V_{\rm eff}=\fr{a_2}{a_3}+(a_1a_2)^{1/4}\fr{\D^2}{\D r_*^2}(a_1a_2)^{-1/4}. 
    \label{Veff}
    \ee
On substituting Eq.~\eqref{p-coeffs}, we have
\begin{align}
    V_{\rm eff}
    =
    f\left[
    \fr{\ell(\ell+1)}{r^2}
    -\fr{3\rH}{r^3} 
    \right.
    \left.
    +\fr{1152(\ell-1)\ell(\ell+1)(\ell+2)}{r^2}\bra{\fr{\rH}{r}}^8\fr{\mathtt{a}(C_{\rm BG})}{(\rH\Lambda)^6}
    \right],
    \label{Veff_ex}
\end{align}
where we have omitted terms of higher order in the higher derivative correction.
If we choose $\mathtt{a}(C) = [C/(12 \Lambda^{4})]^{p}$ so that
    \be
    \mathtt{a}(C_{\rm BG})=\fr{1}{(\rH\Lambda)^{4p}}\bra{\fr{\rH}{r}}^{6p},
    \ee
with $p$ a non-negative integer, then Eq.~\eqref{Veff_ex} reads
    \be
    V_{\rm eff}=f\brb{\fr{\ell(\ell+1)}{r^2}-\fr{3\rH}{r^3}+\fr{1152\eEFT(\ell-1)\ell(\ell+1)(\ell+2)}{r^2}\bra{\fr{\rH}{r}}^{6p+8}}.
    \label{Veff_ex2}
    \ee
Note in passing that, for simplicity, we have used the same symbol~$\Lambda$ for the mass scale in the ansatz for $\mathtt{a}(C)$ and in the action~\eqref{action}, although they can in general be different.
Here, we have defined
    \be
    \eEFT\coloneqq \fr{1}{{(\rH\Lambda)^{4p+6}}},
    \ee
which is assumed to be small.\footnote{Note that $\epsilon_{\rm EFT}$ can, in principle, be either positive or negative.
The key point is that the power of the mass scale~$\Lambda$ simply reflects dimensional counting, and $\Lambda$ itself carries no independent physical meaning.
Therefore, although $\epsilon_{\rm EFT}$ involves $\Lambda^{4p+6}$ with an even power, its sign is not constrained.
Meanwhile, the sign of the higher curvature coupling may be fixed under specific assumptions about the ultraviolet completion of the EFT (see, e.g., Ref.~\cite{Endlich:2017tqa} for a related discussion), although such considerations lie beyond the scope of the present work.}
It should be noted that the above expression of $V_{\rm eff}$ is valid for
    \be
    \ell\ll \eEFT^{-1/2}.
    \ee
Indeed, if $\ell$ is of ${\cal O}(\eEFT^{-1/2})$ or larger, the third term inside the square brackets in Eq.~\eqref{Veff_ex2}, corresponding to the leading-order EFT correction, can be comparable to the first term, which means that the perturbative treatment is no longer valid.

In what follows, we apply the parametrized quasinormal ringdown formalism to our model and examine how the QNMs are modified relative to those in GR.

\section{Parametrized QNM formalism}\label{sec:pQNM}

We briefly review the parametrized quasinormal ringdown formalism~\cite{Cardoso:2019mqo, Hirano:2024fgp}. 
In a broad class of gravity theories, the master equation for the odd-parity perturbations about static and spherically symmetric black hole solutions can be written in the frequency domain as
\begin{align}
    f \frac{\rmd }{\rmd r} \bra{f \frac{\rmd \psi}{\rmd r}} + \bra{\omega^2 - V_{\rm eff}} \psi = 0. 
    \label{eq:parametrizedeq}
\end{align}
The effective potential can be written in the following form:
\begin{align}
    V_{\rm eff} = V_{\rm RW} + \delta V,
\end{align}
where $V_{\rm RW}$ is the Regge-Wheeler potential, 
\begin{align}
    V_{\rm RW} = f \brb{\frac{\ell(\ell+1)}{r^2} - \frac{3r_{\rm H}}{r^3}}.
\end{align}
The deviation from the Regge-Wheeler potential $V_{\rm RW}$ is characterized by $\delta V$, which can be written as a power series in $r_{\rm H}/r$ multiplied by $f(r)$, as follows:
\begin{align}
    \delta V 
    = \sum_{j=0}^{\infty}\alpha_{j} \delta v_{j} 
    \coloneqq \frac{f}{r_{\rm H}^2} \sum_{j=0}^{\infty} \alpha_{j} \bra{\frac{r_{\rm H}}{r}}^{j}. \label{delta_V}
\end{align}
Here, each $\delta v_j$ has its own characteristic radial dependence, $f(r)(r_{\rm H}/r)^j$, and the parameter~$\alpha_j$ controls its magnitude.
The set of parameters~$\{\alpha_j\}$ encodes the deviation from GR, which is assumed to be small and treated perturbatively. As shown in the previous section, for the EFT described by the action~\eqref{action} with $\mathtt{a}(C) = [C/(12 \Lambda^{4})]^{p}$, 
the deviation from the Regge-Wheeler potential takes the form
\begin{align}
    \delta V = \frac{f}{r_{\rm H}^2}
    \brb{1152 \epsilon_{\rm EFT}(\ell-1) \ell (\ell+1)(\ell+2)} \bra{\frac{r_{\rm H}}{r}}^{6p+10}. 
\end{align}
Therefore, this expression fits into the form of the parametrized potential in Eq.~\eqref{delta_V}, where the only nonvanishing coefficient is $\alpha_j$ with $j=6p+10$, given explicitly by
\begin{align}
    \alpha_{j} = 1152 \epsilon_{\rm EFT}(\ell-1) \ell (\ell+1)(\ell+2).
\end{align}
In what follows, instead of $\epsilon_{\rm EFT}$, we use $\alpha_j$ to control the effect of the higher curvature term.

The QNMs are solutions to Eq.~\eqref{eq:parametrizedeq} satisfying both the purely outgoing boundary condition at the spatial infinity $r_{*} \to \infty$ and the purely ingoing boundary condition at the horizon $r_{*} \to -\infty$, where $r_* = r + r_{\rm H} \log \brb{(r-r_{\rm H})/ r_{\rm H}}$ is the tortoise coordinate. 
If the parameters~$\alpha_{j}$ are small and the shifts in the QNM frequencies are perturbative, we can expand the QNM frequencies around their GR values to leading order in $\alpha_j$ as follows:
\begin{align}
    \omega_{n} = \omega^\text{Sch}_{n} + \sum_{j=0}^{\infty} \alpha_{j} e_{j,n},
    \label{eq:QNMfreEFT}
\end{align}
where $\omega^\text{Sch}_{n}$ is the Schwarzschild QNM frequencies with overtone number~$n$. For example, $\omega^\text{Sch}_{0} = (0.7473434 - 0.1779246i) / r_{\rm H}$ for the $\ell=2$ fundamental mode (labeled by $n=0$). 
The numerical constants~$e_{j,n}$ can be determined independently of $\alpha_{j}$, and their numerical values were obtained in Refs.~\cite{Cardoso:2019mqo,Hirano:2024fgp}. 
Once the set of expansion coefficients~$\{\alpha_j\}$ is specified for the setup of interest, the corresponding QNM frequencies can be immediately computed using Eq.~\eqref{eq:QNMfreEFT}.

\begin{figure}[t]
    \includegraphics[width=0.7\columnwidth]{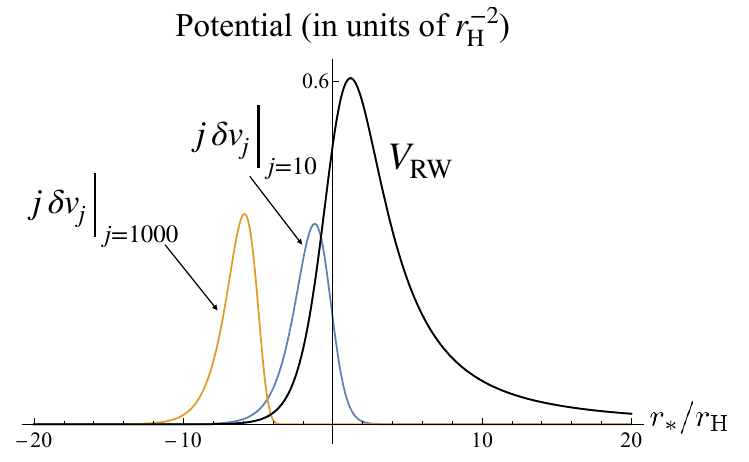}
    \caption{The black line shows the Regge-Wheeler potential~$V_{\rm RW}$, while the blue and orange lines show $j\,\delta v_j$ for $j=10$ and $j=1000$, respectively.
    As the value of $j$ increases, the peak location of 
    $\delta v_j$ approaches to the event horizon.}
    \label{fig:potential}
\end{figure}

We briefly comment on the behavior of the peak location of the correction~$\delta v_{j}$ in the large-$j$ limit. 
In the tortoise coordinate, the peak location of $\delta v_{j}$ for large $j$ is given by
\begin{align}
    r_{*} \approx - r_{\rm H} \log (j/e).
\end{align}
As a result, for large $j$, the correction~$\delta v_{j}$ induces a deformation of the effective potential that is localized near the horizon.
Note in passing that the peak value of $\delta v_j$ is
\begin{align}
    \delta v_{j} \approx \frac{1}{e\,r_{\rm H}^2\,j}.
\end{align}
Figure~\ref{fig:potential} shows the functional form of $\delta v_j$ for $j=10$ and $j=1000$, with $\ell=2$.
Note that we plot $j\,\delta v_j$ rather than $\delta v_j$ itself to make the peak structure more visible. 
The peak location of $\delta v_j$ indeed approaches the horizon as $j$ increases. 

It has been pointed out that a small deformation in the near-horizon region leads to significant changes in the QNM frequencies of the overtones, while the frequency of the fundamental mode is less affected~\cite{Konoplya:2022pbc}. 
This phenomenon can be interpreted in terms of the parametrized QNM formalism as follows. 
As shown in Eq.~\eqref{eq:QNMfreEFT}, the shift of the $n$th overtone frequency from its Schwarzschild value is given by $\alpha_j$ multiplied by $e_{j,n}$, and the asymptotic behavior of $e_{j,n}$ in the large-$j$ limit is given by~\cite{Hirano:2024fgp}
\begin{align}
    e_{j,n}\sim j^{-1 -2 r_{\rm H} \omega_{I n}}, \label{ej_large-j}
\end{align}
where $\omega_{In}\,(<0)$ is the imaginary part of $n$th Schwarzschild QNM frequency, i.e., $\omega^{\text{Sch}}_{n} = \omega_{R n} + i \omega_{I n}$. 
This asymptotic behavior implies that the exponent~$-1-2r_{\rm H} \omega_{I n}$ determines whether $|e_{j,n}|$ diverges or not. 
The exponent is negative for the fundamental mode, while it is positive for the overtones~\cite{Hirano:2024fgp}. 
For example, for $\ell = 2$, one finds $e_{j,0} \sim j^{-0.644}$ for the fundamental mode, whereas $e_{j,1} \sim j^{0.0957}$ for the first overtone. 
It should be noted that this behavior persists even for modes with large $\ell$. 
Accordingly, $e_{j,0}\to 0$ as $j\to \infty$, while $|e_{j,n}|\to \infty$ for all $n\ge 1$.
The tendency of overtone frequencies to be significantly affected by deformations of the effective potential, often referred to as overtone outbursts, has also been observed in the context of spectral instability induced by deformations located far from the black hole~\cite{Barausse:2014tra,Jaramillo:2020tuu,Jaramillo:2021tmt,Cheung:2021bol}.

A caveat is now in order.
For a fixed $j$ and $\alpha_j$, the exponent~$-1-2r_{\rm H} \omega_{I n}$ in Eq.~\eqref{ej_large-j} increases with $n$, implying that the relative QNM shift, $|\alpha_j e_{j,n}/\omega^\text{Sch}_{n}|$, exceeds unity beyond a certain overtone number~$n$.
At that point, the QNM shift can no longer be treated perturbatively.
This indicates that, even if the fundamental mode and the first few overtones remain close to their GR values, sufficiently high overtones inevitably exhibit spectral instability.
This makes the connection to the time-domain analysis subtle. 
Indeed, when the QNM spectrum as a whole (i.e., including the fundamental mode) is significantly modified from the GR case, it has been pointed out that the early-time behavior of the time-domain waveform is typically better described by the vacuum GR QNMs, rather than by the nonperturbatively shifted QNMs~\cite{Kyutoku:2022gbr,Berti:2022xfj}. 
In the next section, however, we perform a fitting analysis using the fundamental mode and the first few overtones, whose deviations from the GR values remain mild, and show that these shifted QNMs can be identified through waveform fitting.

\section{Time-domain waveform and QNM fitting}\label{sec:time-domain}

\subsection{Time-domain integration method}

In order to obtain the time-domain waveform of the perturbation, we use the discretization method developed by Gundlach-Price-Pullin~\cite{Gundlach:1993tp}. 
Note that in this section, we consider the quadrupole mode $\ell = 2$, which is expected to give the dominant contribution.
The equation of motion~\eqref{RWeq} can be written in terms of the null coordinates~$u \coloneqq t - r_{*}$ and $v \coloneqq t + r_{*}$ as follows:
\begin{align}
    -4 \frac{\partial^{2} \Psi}{\partial u \partial v} 
    = 
    V_{\text{eff}} \Psi,
    \label{eq:waveeqinuv}
\end{align}
where we recall that $V_{\rm eff}$ is a function of $r_* = (v-u)/2$. 
We discretize the coordinates~$u$ and $v$ by a uniform grid with spacing~$h$. 
Then, Eq.~\eqref{eq:waveeqinuv} can be discretized as follows:
\begin{align}
    \Psi({\rm N})
    = 
    \Psi({\rm W}) + \Psi({\rm E}) - \Psi({\rm S}) 
    - 
    \frac{h^{2}}{8}V_{\text{eff}}
    ({\rm S})\brb{\Psi({\rm W}) + \Psi({\rm E})} + \mathcal{O}(h^{4}),
    \label{eq:discritized_scheme}
\end{align}
where we have introduced shorthand notations~${\rm S} \coloneqq (u,v)$, ${\rm W}  \coloneqq (u+h, v)$, ${\rm E} \coloneqq (u, v+h)$, and ${\rm N} \coloneqq (u+h, v+h)$.
In our numerical calculation, we set $h=0.0025r_{\rm H}$.

We impose an initial condition on a constant-$t$ surface~$\Sigma$ (i.e., a surface of constant $u+v$). 
We consider a Gaussian wave packet as the initial condition: 
\begin{align}
    \Psi |_{\Sigma} \coloneqq \Psi(2t_{\rm ini}-v,v) = e^{-\frac{1}{2} \bra{\frac{v-v_{0}}{\sigma}}^{2}},
    \label{eq:initialgaussian}
\end{align}
where $t_{\rm ini}$ is the time on the initial surface $\Sigma$, while $v_{0}$ and $\sigma$ are the peak location and the width of the Gaussian wave packet, respectively. 
We choose $t_{\rm ini}=0$ and set the ranges of $u$ and $v$ as $u \in [-320r_{\rm H},200r_{\rm H}]$ and $v \in [-200r_{\rm H},320r_{\rm H}]$. 
We also choose $\sigma = 0.05r_{\rm H}$ and set $v_0$ such that the peak of the initial Gaussian wave packet (in terms of $r_*$) coincides with that of $\delta v_j$. 
In addition, we consider the static initial condition, i.e., $\partial_{t} \Psi |_{\Sigma} =(\partial_{u}+\partial_{v}) \Psi |_{\Sigma} = 0$.
This means that, when the points~${\rm W}$ and ${\rm E}$ lie on the initial surface~$\Sigma$, 
\begin{align}
    \Psi({\rm N}) = \Psi({\rm S}) + \mathcal{O}(h^{4}).
    \label{eq:initialsurface_condition}
\end{align}
Combining Eqs.~\eqref{eq:discritized_scheme} and \eqref{eq:initialsurface_condition}, we have
\begin{align}
    \Psi({\rm N}) 
    = 
    \frac{1}{2} \brb{\Psi({\rm W}) + \Psi({\rm E})}
    - \frac{h^{2}}{16}V_{\text{eff}}
    ({\rm S})\brb{\Psi({\rm W}) + \Psi({\rm E})}
    + \mathcal{O}(h^{4}),
    \label{eq:initialdeltdescritaize}
\end{align}
when ${\rm W,E}\in \Sigma$.\footnote{
Although the point~${\rm S}$ lies in the past of the initial surface~$\Sigma$, the effective potential $V_{\rm eff}$ can be evaluated at ${\rm S}$ using the explicit functional form given in Eq.~\eqref{Veff_ex2}.
}
Then, by using Eqs.~\eqref{eq:discritized_scheme}, \eqref{eq:initialgaussian}, and \eqref{eq:initialdeltdescritaize}, we can obtain the solution~$\Psi(u,v)$ in the numerical domain. 
As shown above, the local discretization error of this method is of $\mathcal{O}(h^4)$.
Therefore, for a fixed numerical domain in which the number of grid points scales as $\mathcal{O}(h^{-2})$, the global error is of $\mathcal{O}(h^2)$. 

Note that we need to fix the values of $j$ and $\alpha_{j}$ to calculate the time-domain waveform. 
We compute the time-domain waveform for three cases, $(j,\alpha_{j})=(10,0.01)$, $(100,0.1)$, and $(1000,0.5)$. 
In what follows, we use a shorthand notation, e.g., $\alpha_{10}=0.01$ to denote the case~$(j,\alpha_{j})=(10,0.01)$. 
These parameter sets are chosen such that the peak height of the correction~$\delta V$ to the potential is of the same order in all cases.

\begin{figure}[t]
    \centering
    \includegraphics[width=\linewidth]{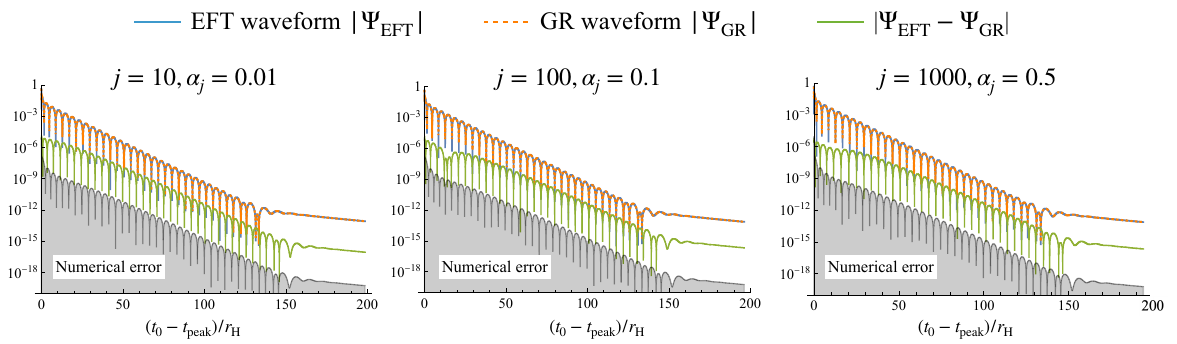}
    \caption{
    The blue solid lines are the time-domain waveforms in the EFTs with $\alpha_{10}=0.01$ (left), $\alpha_{100}=0.1$ (center), and $\alpha_{1000}=0.5$ (right), respectively, while the orange dashed lines are those in GR. 
    Although the EFT and the GR waveforms appear nearly overlapping, the difference between those waveforms (green solid line) are larger than the numerical error (gray solid line). 
    The numerical error is estimated by computing the discretization error from the difference between results obtained with different grid resolutions.
    }
    \label{fig:waveform_all}
\end{figure}
In Fig.~\ref{fig:waveform_all}, the blue solid lines show the time-domain waveforms in the EFTs measured by a static observer located at $r_{*} = 60r_{\rm H}$. 
The left, center, and right panels correspond to the EFTs with $\alpha_{10}=0.01$, $\alpha_{100}=0.1$, and $\alpha_{1000}=0.5$, respectively.
Also, in each case, we show the time-domain waveform in GR (computed using the same initial condition as in the corresponding EFT case) by the orange dashed line. 
We note that the absolute differences between the EFT waveform and GR waveform (green solid lines) are larger than the numerical errors (gray solid lines). 
The latter are estimated by evaluating the discretization error from the difference between results obtained with different grid resolutions (see Appendix~\ref{app:GRfitting} for a more detailed discussion). 
The EFT and GR waveforms appear nearly overlapping in these plots. 
Nevertheless, they differ sufficiently to be distinguishable through waveform fitting, as we discuss below.

\subsection{QNM fitting}

In this subsection, we perform the fitting of the time-domain waveform with templates constructed by the superposition of the damped sinusoids. 
In the present paper, we adopt a fitting approach in which the QNM frequencies are fixed to their theoretical values.
An alternative strategy is the so-called theory-agnostic fit~(see, e.g., Ref.~\cite{Baibhav:2023clw}), where the frequencies are treated as free parameters to be determined from the data.
Fixing the frequencies allows for a direct comparison between the fitted waveform and theoretical predictions for the QNM spectrum.
Therefore, we consider a frequency fixed fitting model~$\psi_{N}$ as follows: 
\begin{align}
    \psi_{N}(t) = {\rm Re} \bra{ \sum_{n = 0}^{N}  A_{n} e^{-i \brb{ \omega_{n} (t-t_{\rm peak}) + \phi_{n}}}},
    \quad 
    t \in [t_{0}, t_{\rm end}],
    \label{eq:EFTtemplate}
\end{align}
where the frequencies~$\omega_{n}$ are fixed with Eq.~\eqref{eq:QNMfreEFT}. 
We refer to the fit using QNM frequencies $\omega_{n}$ with non-zero $\alpha_{j}$ as the EFT fit, and to the fit with $\omega_{n} = \omega_{n}^{\text{Sch}}$ (i.e., $\alpha_{j} = 0$) as the GR fit. 
In the fitting model, $t_{0}$ and $t_{\rm end}$ are respectively the start time of the fitting and the end time of the time-domain waveform.
Here, $N$ denotes the number of overtones included in the template; including the fundamental mode, the total number of modes is therefore $N+1$. 
We set $t_{\rm end} = 100r_{\rm H}$ for all analyses. 
We have confirmed that varying $t_{\rm end}$ does not cause significant changes for the results. 
The fitting model contains $2N+2$ fitting parameters~$A_{n}$ and $\phi_{n}$, and we use the $\mathtt{Mathematica}$ function~$\mathtt{NonlinearModelFit}$ to find their best-fit values.  
Note that the start time of the fitting is chosen to be after the peak time~$t_{\rm peak}$ of the time-domain waveform, i.e., $t_0 \ge t_{\rm peak}$. 

\begin{figure}[t]
    \centering
    \includegraphics[width=0.8\linewidth]{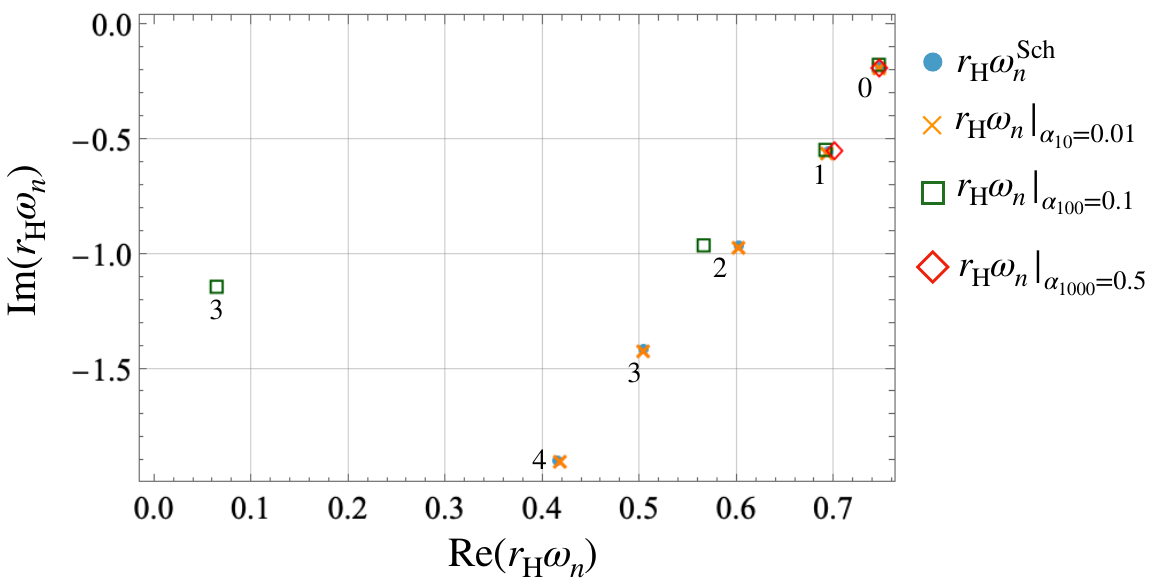}
    \caption{QNM frequencies for the three representative cases, $\alpha_{10}=0.01$, $\alpha_{100}=0.1$, and $\alpha_{1000}=0.5$, together with the GR values.
    The vacuum GR QNMs are shown as blue dots, while the QNMs for $\alpha_{10}=0.01$, $\alpha_{100}=0.1$, and $\alpha_{1000}=0.5$ are indicated by orange crosses, green squares, and red diamonds, respectively.
    The numbers attached to the data points denote the overtone number~$n$.
    We show only those QNMs whose deviations from the GR values remain perturbative.}
    \label{fig:omegaplane}
\end{figure}
As mentioned earlier, we consider the three representative cases, $\alpha_{10}=0.01$, $\alpha_{100}=0.1$, and $\alpha_{1000}=0.5$.
Figure~\ref{fig:omegaplane} shows the QNM frequencies for these three cases, together with the GR values. 
The QNM frequencies were obtained using the formula~\eqref{eq:QNMfreEFT}, with the numerical values of $e_{j,n}$ summarized in Appendix~\ref{app:ejinlargej}. 
The vacuum GR QNMs are shown as blue dots, while the QNMs for $\alpha_{10}=0.01$, $\alpha_{100}=0.1$, and $\alpha_{1000}=0.5$ are indicated by orange crosses, green squares, and red diamonds, respectively. 
Note that the figure displays only those QNMs whose deviations from the GR values remain perturbative, i.e., those satisfying $|\alpha_{j} e_{j,n} / \omega^{\rm Sch}_{n}| < 1$. 
For $\alpha_{10}=0.01$, the QNMs up to the fourth overtone lie within the perturbative regime.
By contrast, the perturbative treatment is valid only up to the third overtone for $\alpha_{100}=0.1$, and only up to the first overtone for $\alpha_{1000}=0.5$.
In our fitting analysis, we therefore take into account only the QNMs shown in Fig.~\ref{fig:omegaplane}. 
For $n=3$ in the case of $\alpha_{100}=0.1$, we find $|\alpha_{j}e_{j,n} / \omega^{\rm Sch}_{n}|\simeq 0.34$, indicating that the perturbative treatment is approaching the edge of its validity.

Let us also explain how to quantify the goodness of the fit.
A common approach is to use the mismatch defined by 
\begin{align}
    \mathcal{M} 
    \coloneqq 
    1 - \frac{\langle \Psi | \psi_{N} \rangle }{\sqrt{ \langle \Psi |  \Psi \rangle  \langle \psi_{N} | \psi_{N} \rangle }},
\end{align}
where the scalar product~$\langle F | G \rangle$ is defined as 
\begin{align}
    \langle F | G \rangle
    \coloneqq 
    \int_{t_{0}} ^{t_{\rm end}} F(t) G (t) \, \rmd t, 
\end{align}
for real functions $F$ and $G$. 
Then, one has $0 \le \mathcal{M} \le 1$, and $\mathcal{M}=0$ if and only if $\Psi\propto \psi_N$. 
In general, a small mismatch indicates that the data can be well described by the fitting model. 
However, a small mismatch alone is not sufficient evidence of a good fit in time-domain ringdown analyses~\cite{Baibhav:2023clw,Nee:2023osy,Takahashi:2023tkb}. 
In fact, a small mismatch can be obtained simply by including more overtones in the fitting model, even if this merely reflects overfitting rather than a genuinely improved fit.
In order to confirm that overfitting does not occur, we shall also check the stability of the fitting parameters~$A_{n}$ and $\phi_{n}$ when varying $t_{0}$.

\subsection{Results}\label{ssec:results}

\begin{figure}[t]
    \includegraphics[width=0.92\textwidth]{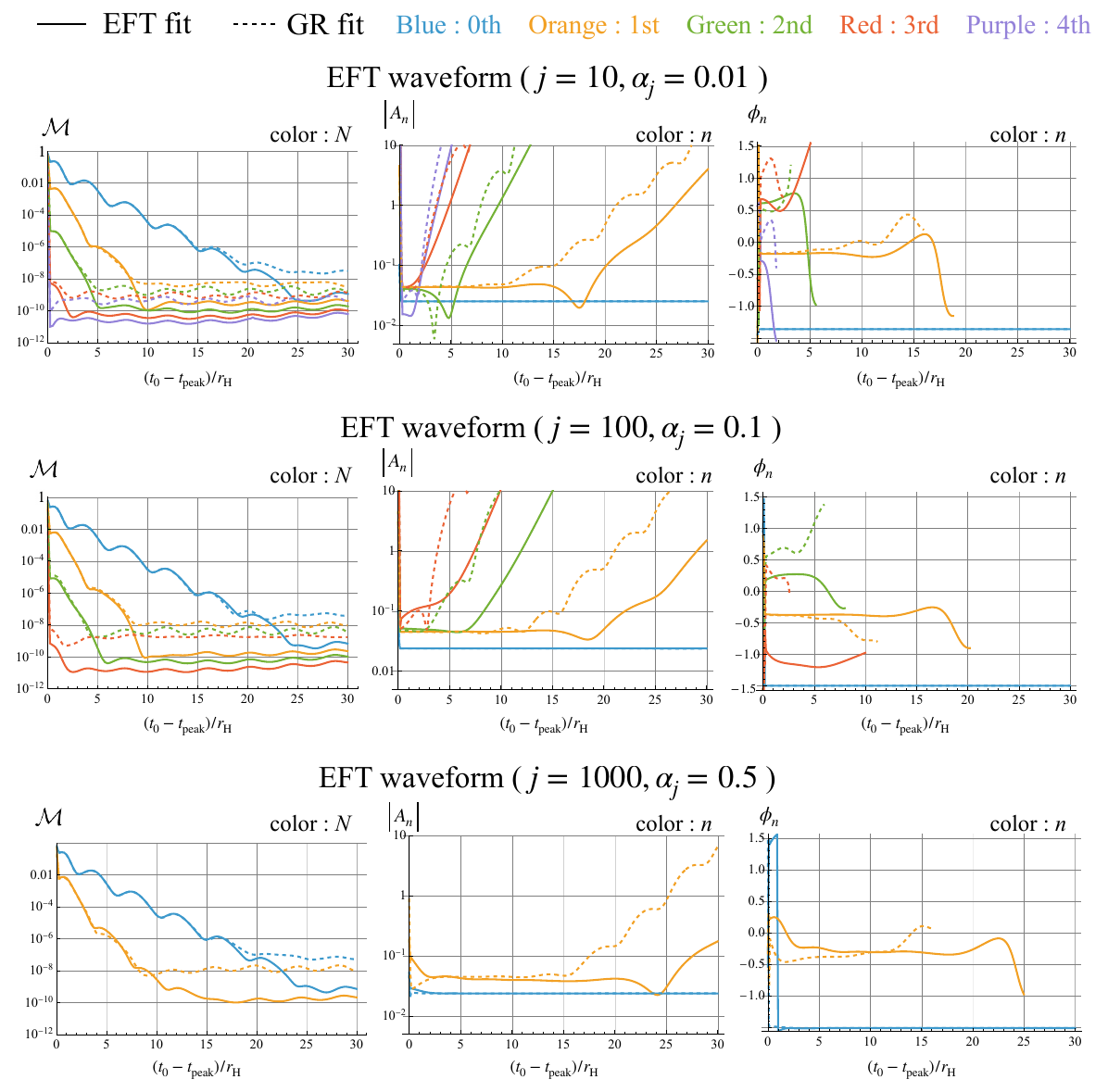}
    \caption{The mismatch~$\mathcal{M}$, the amplitudes~$A_{n}$, and the phases~$\phi_{n}$ as functions of $(t_{0} - t_{\rm peak})/r_{\rm H}$ for the three representative cases, $(j, \alpha_{j}) = (10,0.01)$, $(100, 0.1)$, and $(1000,0.5)$. 
    In the mismatch plots, the colors correspond to the fitting model~\eqref{eq:EFTtemplate} with different values of $N$. 
    We use up to $N=4$ for $\alpha_{10}=0.01$, up to $N=3$ for $\alpha_{100}=0.1$, and up to $N=1$ for $\alpha_{1000}=0.5$. 
    In the plots of $A_n$ and $\phi_n$, the value of $N$ is fixed to its maximal value for each case, and the colors correspond to different values of $n$.
    The solid lines are the results for the EFT fit, while the dashed lines are those for the GR fit. 
    }
    \label{fig:allfigs_M_A_TH}
\end{figure}

Figure~\ref{fig:allfigs_M_A_TH} shows the mismatch~$\mathcal{M}$, the amplitudes~$A_{n}$, and the phases~$\phi_{n}$ as functions of $(t_{0} - t_{\rm peak})/r_{\rm H}$ for the three representative cases~$(j, \alpha_{j}) = (10,0.01)$, $(100, 0.1)$, and $(1000,0.5)$. 

For these three cases, the qualitative behavior of these quantities is essentially the same. 
In both the EFT and GR fits, as overtones are added to the fitting model, the minimum of the mismatch decreases, and the fitting start time~$t_{0}$ at which the mismatch takes its minimum moves toward the peak time~$t_{\text{peak}}$. 
On the other hand, there are notable quantitative differences between the EFT fit and the GR fit. 
We find that the minimum of the mismatch obtained with the EFT fit is smaller by one or two orders of magnitude than that obtained with the GR fit, and this trend does not depend sensitively on the number of modes included in the fitting model. 
An advantage to take into account overtones in the fitting model is that the difference in the mismatch between the EFT and GR fits becomes apparent at earlier times.\footnote{
The authors of Ref.~\cite{Konoplya:2022pbc} pointed out that the overtone effects can cause differences in time-domain waveforms for test fields in the early-time regime.} 
For the mismatch including only the fundamental mode, the difference between the EFT and GR fits starts to appear at $t_{0} - t_{\rm peak} \simeq 15r_{\rm H}$ in all cases. In contrast, when the fitting model includes up to the first overtone, the difference emerges much earlier, at $t_{0} - t_{\rm peak} \simeq 7r_{\rm H}$, again consistently across all cases. 
These findings suggest that the EFT overtones may play an important role when analyzing actual observational data, especially in the early stage of the ringdown where the signal-to-noise ratio is high and the waveform is most sensitive to deviations from GR. 

The stability of the fitting parameters with respect to the fitting start time $t_0$ reveals a similar trend across all cases. 
For all parameter sets, the fundamental-mode amplitude $A_0$ and phase $\phi_0$ remain nearly constant for both the EFT and GR fits. 
For overtones, the duration over which $A_n$ and $\phi_n$ stay approximately constant becomes shorter, because overtone contributions decay rapidly and dominate only during the short initial stage of the waveform. 
Nevertheless, in every case, the EFT fits maintain parameter stability over a longer interval of $t_0$ than the GR fits, reinforcing that the EFT QNMs more accurately capture the dynamical content of the waveform. 

Let us also comment on the behavior of the time-domain waveforms in relation to the shifted QNMs shown in Fig.~\ref{fig:omegaplane}. 
For $n=3$ in the case of $\alpha_{100}=0.1$, the perturbative treatment is approaching the edge of its validity, as indicated by $|\alpha_{j}e_{j,n} / \omega^{\rm Sch}_{n}|\simeq 0.34$. 
However, even in this case, the EFT fit still outperforms the GR fit, as shown in Fig.~\ref{fig:allfigs_M_A_TH}. 
Aside from this borderline case, the shifted QNM frequencies remain close to the vacuum GR QNM frequencies in the complex frequency plane.
It is intriguing to note that, even when the QNM frequencies deviate only slightly from their GR values, the EFT QNMs still provide a better fit to the time-domain waveform than the GR QNMs, as shown in Fig.~\ref{fig:allfigs_M_A_TH}.
This indicates that even small shifts in the QNM frequencies can affect the waveform fitting. 

A related observation was reported in Ref.~\cite{Rosato:2024arw}, where a time-domain waveform was compared with superpositions of the GR QNMs and shifted QNMs.
A key difference, however, lies in the nature of the QNM modifications: in Ref.~\cite{Rosato:2024arw}, the entire QNM spectrum exhibits sizable deviations from the GR values, whereas in our analysis we find that even mild shifts of the QNM frequencies are sufficient to reduce the mismatch and improve the fit.

A related but conceptually distinct approach was explored in Ref.~\cite{Oshita:2025ibu}, where the authors argued that shifted QNMs caused by a bump located far from the peak of the effective potential can form a better ``basis'' for describing the full waveform, including not only the damped oscillatory ringdown but also the late-time power-law tail.
When the bump is located far from the potential peak, a large number of shifted QNMs accumulate densely near the real axis, which has been shown to be a generic feature of spectral instability in bump-type models~\cite{Motohashi:2024fwt,Ianniccari:2024ysv}.
A decomposition in terms of such shifted QNMs then results in a representation that closely resembles a Fourier decomposition.
While such an approach may provide an apparently good description of the waveform, its connection to the physical interpretation of QNMs as characteristic excitations of the black hole spacetime is less transparent.
In contrast, our analysis focuses on the regime where QNMs retain their interpretation as damped oscillatory modes of the black hole spacetime, and demonstrates that even small shifts in the QNM spectrum can have a measurable impact on waveform fitting.

\section{Summary and discussions}\label{sec:summary}

In the present paper, we have studied linear perturbations around Schwarzschild black holes in higher curvature gravity. 
In particular, we have focused on how near-horizon deformations of the effective potential originating from higher curvature corrections affect the QNM spectrum and the time-domain waveform.
Using the parametrized QNM formalism, we have computed the shift of the QNMs up to first order in the deviation from GR. As pointed out in Ref.~\cite{Hirano:2024fgp}, as the order of the higher curvature term increases, the overtone frequencies 
tend to deviate progressively from their GR values, while the fundamental mode is less affected. 
This behavior suggests a close connection between the higher curvature corrections and the overtone outburst, a phenomenon known in the context of spectral instability. 
In typical cases
of spectral instability, when the entire QNM spectrum significantly deviates from the GR values, the early-time behavior of the time-domain waveform is still often
better described by the vacuum GR QNMs.
In contrast, we have considered a situation in which the deviations of the QNM frequencies from their GR values remain mild for the fundamental mode and the first few overtones, while the higher overtones exhibit non-perturbative deviations. 
In this regime, it is not a priori obvious how such frequency shifts manifest themselves in the time-domain waveform. Nevertheless, we have found that the waveform can be well described by a superposition of the mildly shifted QNMs.

To quantitatively assess the impact of these QNM frequency shifts on observable signals, 
we have performed ringdown fits to the time-domain waveforms in our EFT with a higher-curvature term, using the EFT and GR QNM spectra. 
When overtones are included in the fitting model, differences between the EFT and GR fits 
emerge at earlier time. 
For all parameter choices, the EFT fit yields
a smaller mismatch than the GR fit, demonstrating that the EFT QNMs more accurately capture the waveform. 
These features indicate that the early-time stage of the ringdown, which is dominated by the overtone contributions, can enhance sensitivity to the near-horizon physics. 

We have also examined the stability of the best-fit amplitudes and phases with respect to the fitting start time. 
While in GR fit the overtone parameters are stable only over a short initial interval due to a rapid decay of overtone contributions, the EFT fit maintains stable values over a longer 
range of fitting start times than the GR fit. 
Even in regimes where the order of higher curvature term is large, the EFT QNMs are excited and leave imprints on the time-domain waveform that become identifiable through ringdown fitting.

Our results show that ringdown observations, particularly in the early-time stage and with overtone information included, can serve as a sensitive probe of near-horizon modifications predicted by higher curvature gravity. 
Extending this analysis to rotating black holes and developing data-analysis strategies optimized for the early ringdown constitute promising directions for future work.

\begin{acknowledgments}
We would like to thank Tsutomu Kobayashi and Karim Noui for useful comments and discussions. This work was supported in part by JSPS KAKENHI Grant Numbers JP22K03626 (M.K.), JP22K03639 (H.M.), and JP23K13101 (K.T.). 
\end{acknowledgments}

\appendix

\section{Fitting the GR waveform with GR QNMs}
\label{app:GRfitting}

\begin{figure}[t]
    \includegraphics[width=0.6\textwidth]{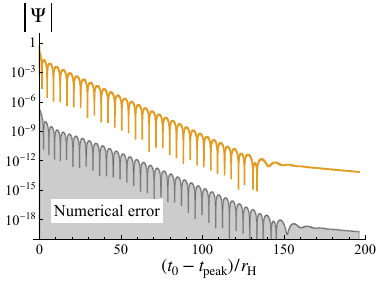}
    \caption{
    The orange solid line is the time-domain waveform in GR, while the gray solid line is the numerical error. 
    The dominant contribution to the numerical error comes from the discretization error, which is estimated by comparing results obtained with different resolutions. 
    }
    \label{fig:waveformGR}
\end{figure}

In this appendix, we show the results of the GR waveform fitting with the GR QNMs as a consistency check. 
This analysis serves to confirm the robustness of our fitting procedure and to provide a reference case for comparison with the EFT waveforms analyzed in the main text. 

In Fig.~\ref{fig:waveformGR}, we show the time-domain waveform in GR as the orange solid line. 
The initial Gaussian wave packet has the width $\sigma = 0.05 r_{\rm H}$, and the observer is located at $r_{*} = 60 r_{\rm H}$. 
These parameters are also used in the calculation of the time-domain waveform in the EFT in the main text. 
The numerical error is displayed by the gray solid line in Fig.~\ref{fig:waveformGR}. 
We have confirmed that the discretization error dominates over other sources of numerical uncertainty, such as the round-off error.
In our numerical method, the (global) discretization error scales as $\mathcal{O}(h^2)$ [see the discussion below Eq.~\eqref{eq:initialdeltdescritaize}], and it can be estimated as follows.
Let $\Psi_h(t)$ denote the waveform computed with grid spacing~$h$ as observed by a static observer, and let $\Psi_0(t)$ denote the exact solution obtained in the limit $h\to0$.
We then have
    \begin{align}
    \Psi_h(t)=\Psi_0(t)+\Delta(t)h^2+\mathcal{O}(h^3),
    \end{align}
where $\Delta(t)$ is a function of time.
Accordingly, the difference between the waveforms computed with grid spacings~$h$ and $2h$ is, at leading order in $h$,
    \begin{align}
    \Psi_{2h}(t)-\Psi_h(t)\simeq 3\Delta(t)h^2.
    \end{align}
We have verified this relation for several values of $h\le 0.02r_{\rm H}$.
Therefore, the discretization error can be estimated as
    \begin{align}
    \Psi_h(t)-\Psi_0(t)\simeq \frac{1}{3}\brb{\Psi_{2h}(t)-\Psi_h(t)}.
    \end{align}
The numerical error shown in Fig.~\ref{fig:waveformGR} is obtained using this procedure.
Note that we perform the numerical calculation with the grid spacing $h = 0.0025 r_{\rm H}$.

\begin{figure}[t]
    \includegraphics[width=\textwidth]{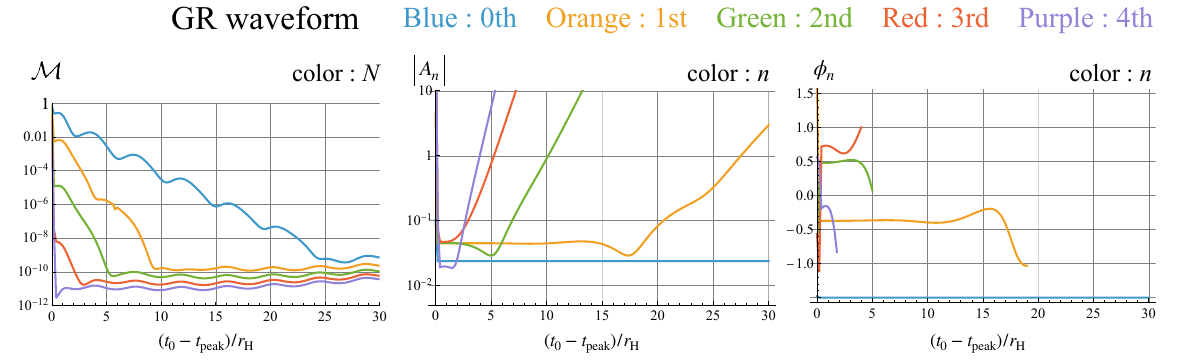}
    \caption{Mismatch~$\mathcal{M}$, amplitudes~$A_n$, and phases~$\phi_n$ for the GR waveform fitted with the GR QNM spectrum. The same color scheme as in Fig.~\ref{fig:allfigs_M_A_TH} is used in each plot. }
    \label{fig:GR_allfigs}
\end{figure}

Figure~\ref{fig:GR_allfigs} show the mismatch~$\mathcal{M}$, the amplitudes~$A_{n}$, and the phases~$\phi_{n}$. 
As in the EFT case studied in the main text, the mismatch decreases monotonically when higher overtones are included in the fitting model. 
The minimum of the mismatch is achieved at earlier times when the overtones are taken into account, and this result is consistent with the fact that the early-time waveform is dominated by overtone contributions. 
The stability of the best-fit amplitudes~$A_n$ and phases~$\phi_n$ also shows the same qualitative pattern as in the EFT case discussed in the main text. For the fundamental mode, both $A_0$ and $\phi_0$ remain nearly constant over a wide range of fitting start times. 
The overtones exhibit stability only over a shorter interval near the peak of the waveform, as their contributions decay rapidly and are dominant only during the early stage of the waveform. 

As a consistency check on the numerical accuracy of our analysis, we examined how accurately the QNM frequencies can be extracted from the GR waveform. 
To do so, we introduced a small offset in the fitting model by replacing the GR frequency with $\omega_{n}^{\mathrm{Sch}} + \delta\omega_{n}$ and evaluated 
the value of $\delta\omega_{n}$ that provides the smallest mismatch. 
If the time-domain waveform were completely described by a superposition of the GR QNMs, $\delta \omega_{n}$ that minimizes the mismatch would be zero. 
In practice, however, the time-domain waveform contains contributions which cannot be described by the QNMs, with the dominant such contribution arising from the late-time tail. 
Therefore, the offset~$\delta \omega_{n}$ that provides the smallest mismatch is generally nonzero. 
Evaluating this optimal $\delta \omega_{n}$ allows us to estimate how accurately the QNM frequencies can be extracted from the waveform. 
We found that, for the fundamental mode as well as the first and second overtones, the GR QNM frequencies can be recovered from our numerical waveform with an accuracy of approximately five to six significant digits. 

\section{\texorpdfstring{Numerical values of $e_{j,n}$}{Numerical values of ej}}
\label{app:ejinlargej}

\begin{table}[t]
    \caption{QNM frequencies in GR ($\ell=2$)}
    \label{table:omegasch}
    \centering
    \begin{tabular}{cc}
      \hline
      $n$ & $r_{\rm H}\omega^{\rm Sch}_{n}$   \\
      \hline 
      0 & $0.747343368836084 - 0.177924631377871 i$   \\
      1 & $0.693421993758327 - 0.547829750582470 i$   \\
      2 & $0.602106909224733 - 0.956553966446144 i$   \\
      3 & $0.50300992437118 - 1.41029640486699 i$   \\
      4 & $0.41502915962613 - 1.89368978173270 i$   \\
      \hline
    \end{tabular}
    \caption{$e_{j,n}$ for $j=10$ ($\ell=2$)}
    \label{table:ejj10ell2}
    \centering
    \begin{tabular}{cc}
      \hline
      $n$ & $r_{\rm H}e_{10,n}$   \\
      \hline 
      0 & $0.00368531438581036 + 0.00652444451724573 i$   \\
      1 & $0.0144010839340230 + 0.0233065697436193 i$   \\
      2 & $0.0480745969771377 + 0.0522805954645665i$   \\
      3 & $0.134865872349529 + 0.105123598889745i$   \\
      4 & $0.341401447538711 + 0.202009507605051i$   \\
      \hline
    \end{tabular}
    
    \caption{$e_{j,n}$ for $j=100$ ($\ell=2$)}
    \label{table:ejj100ell2}
    \centering
    \begin{tabular}{cc}
      \hline
      $n$ & $r_{\rm H}e_{100,n}$   \\
      \hline 
      0 & $6.43382090530 \times 10^{-7} - 0.001138507981692972i$   \\
      1 & $-0.0140538912436019 - 0.0191292171076602i$   \\
      2 & $-0.353349774328950 - 0.061498907374063 i$   \\
      3 & $-4.37901117948735 + 2.65601991377345i$   \\
      4 & $-30.1324771895834 + 62.7236815904998 i$ \\
      \hline
    \end{tabular}

      \caption{$e_{j,n}$ for $j=1000$ ($\ell=2$)}
      \label{table:ejj1000ell2}
    \begin{tabular}{cc}
      \hline
      $n$ & $r_{\rm H}e_{1000,n}$   \\
      \hline 
      0 & $-0.000079131144903428 + 0.000238572496454560i$   \\
      1 & $0.0160233288654551 + 0.0238482241305253i$   \\
      2 & $2.71955972866960 - 0.63875652583774 i$ \\
      3 & $45.814519689636 - 305.730976514491 i$ \\
      4 & $-27258.4815735064 - 24194.2706613034 i$ \\
      \hline
    \end{tabular}
\end{table}

In this appendix, we briefly summarize the numerical values of $e_{j,n}$ used in our calculations.
We use the fact that the coefficients~$e_{j,n}$ for odd-parity perturbations satisfy the following recurrence relation~\cite{Kimura:2020mrh,Hatsuda:2023geo,Hirano:2024fgp}:
\begin{align}
    c_{j+1}e_{j+1,n} + c_{j+3}e_{j+3,n} + c_{j+4}e_{j+4,n} + c_{j+5}e_{j+5,n} = 0\quad
    (j\ge -1),
    \label{eq:recursionrelation}
\end{align}
where the coefficients are expressed as follows:
\begin{align}
\begin{split}
    c_{j+1}
    &=
    -4 j\bra{r_{\rm H}\omega^{\rm Sch}_n}^{2},
    \\
    c_{j+3}
    &=
    - (j+1)\brb{-4\ell(\ell+1)+j(j+2)},
    \\
    c_{j+4}
    &=
    (2j+3)\brb{-6 -2\ell(\ell+1) +j(j+3)},
    \\
    c_{j+5}
    &=
    - (j-2)(j+2)(j+6).
\end{split}
\end{align}
Using the recurrence relation~\eqref{eq:recursionrelation}, one can express $e_{j,n}$ for $j\ge 3$ in terms of $e_{0,n}$, $e_{2,n}$, and $e_{7,n}$, whose numerical values can be found in Refs.~\cite{Kimura:2020mrh,Hirano:2024fgp}.
Note in passing that $e_{1,n}$ does not appear explicitly in the recurrence relation and therefore does not need to be specified.
The QNM frequencies in GR for $\ell=2$, which are necessary to fix the coefficient~$c_{j+1}$, are shown in Table~\ref{table:omegasch} up to the fourth overtone.\footnote{High-precision datasets of QNM frequencies for both Schwarzschild and Kerr black holes can be found in Ref.~\cite{motohashi_2025_14380191}.}
The numerical values of $e_{10,n}$, $e_{100,n}$, and $e_{1000,n}$ up to the fourth overtone, computed using the recurrence relation~\eqref{eq:recursionrelation}, are shown in Tables~\ref{table:ejj10ell2}, \ref{table:ejj100ell2}, and \ref{table:ejj1000ell2}, respectively.
As discussed in the main text, the QNM shifts remain perturbative for $0\le n\le 4$ when $\alpha_{10}=0.01$, for $0\le n\le 3$ when $\alpha_{100}=0.1$, and for $0\le n\le 1$ when $\alpha_{1000}=0.5$.

\bibliographystyle{JHEPmod}
\bibliography{apssamp.bib}
\end{document}